\documentclass[11pt]{revtex4}
\usepackage{hyperref}
\begin{document}

\title{Spontaneous breaking of nilpotent symmetry in boundary BLG theory}

 \author{ Sudhaker Upadhyay}
 \email {  sudhakerupadhyay@gmail.com; 
 sudhaker@iitk.ac.in}

\affiliation {Department of Physics, Indian Institute of Technology Kanpur, Kanpur 208016, India }

\begin{abstract}
We exploit boundary term to preserve the supersymmetric gauge invariance of Bagger--Lambert--Gustavsson (BLG) theory.
The fermionic rigid BRST and anti-BRST symmetries are studied in linear and non-linear gauges.
Remarkably, for Delbourgo-Jarvis-Baulieu-Thierry-Mieg (DJBTM) type  gauge   the spontaneous breaking of BRST symmetry occurs  in  the BLG theory.
The responsible guy for such spontaneous breaking is  
ghost-anti-ghost condensation. Further, 
we discuss the ghost-anti-ghost 
condensates  in the modified maximally Abelian (MMA) gauge  in  the BLG theory.   
  
 \end{abstract}

\maketitle 
\section{Introduction} 
 
 Bagger and Lambert \cite{bl0,bl00, bl01} and Gustavsson \cite{bl02} proposed a formalism 
 to describe  multiple M2-branes and it was found that the generalized Jacobi identity for Lie
3-algebra  that generalizes the notion of the Lie algebra is essential to define the action with ${\cal N}= 8$ supersymmetry. 
The 3-algebras,
and, in general, n-algebras were introduced by V. Filippov \cite{fil} however, they were intimately 
related to the Nambu bracket \cite{nam}.
A coupling of the BLG theory   (a three dimensional field theory)  to $D = 3$, ${\cal N} = 8$ conformal supergravity is invoked in \cite{jh,jh1}. In fact, the conformal supergravity multiplet considered there in the
component field formulation contains a dreibein, eight Rarita-Schwinger fields and $SO(8)$
gauge fields. The local symmetry transformation laws of these fields   and an
invariant Lagrangian of the coupled theory have been constructed. The Lagrangian contains 
  the conformal supergravity multiplet, and commutators of local on-shell super-transformations.

 So far the only
explicit example of Lie 3-algebra ever considered for the BLG model is ${\cal A}_4$ which is the
 $SO(4)$-invariant algebra with 4 generators.  For a more concrete understanding of  such
model, we need to study more explicit examples of Lie 3-algebra.   
With the advent of the  BLG  model, an alternative
method to construct a 5--brane action   with the gauge
symmetry associated with the group
of volume preserving diffeomorphisms was proposed in \cite{ber1,ber2,ber4}. 
Although some steps
have already been undertaken \cite{pas,pas1}, the equivalence of
the BLG model to the M5--brane description reported in  \cite{ban,ban1} is still to be verified.   
The BLG theory 
 has been used for analysing a system of M2-branes ending on a M5-brane, so  it is
worthwhile to study the BLG theory in presence of a boundary.
 The supersymmetric theories   in presence of a boundary have been studied extensively \cite{ld,fa}.
However, the multiple  branes  with a boundary are   studied in \cite{ds,mf,mf1}.
It is well-known that in a supersymmetric theory, the presence of a boundary  breaks  the supersymmetry.  This is because the 
boundary
obviously breaks translational symmetry and since supersymmetry closes on translations.
Incidentally, it has been found that the boundary BLG theory respects the gauge symmetry \cite{mir}.  
The BRST Symmetry for the theory of M2-branes are investigated in \cite{mi,msb,msb1,msb2}.

On the other hand, the   ghost-anti-ghost condensation present in the  theories
possessing gauge symmetry  play
an important role
 \cite{kon, z1, z2, z3, z4}. 
In this framework, it was shown that such condensation  leads   a spontaneous breaking  of
supersymmetries present in the theory. 
In fact, for non-Abelian theories in   the maximally Abelian (MA) gauge  the ghost-anti-ghost
condensation offers  
a mechanism to provide the masses of off-diagonal gluons and off-diagonal ghosts
 \cite{sch, kon1}. This mechanism  gets relevance in
infrared Abelian dominance \cite{hooft},  
which justifies the dual
superconductor picture \cite{nambu, mandal, poly} of QCD vacuum for explaining quark confinement \cite{kon1,kon2,kon3,kon4}. 
The breaking of spontaneous breaking  of
supersymmetries  have led  many interesting consequences in different
situations \cite{kon, n0, n1, n2, n4}.
Recently, the presence of  ghost-anti-ghost condensation  has been occurred  in ABJM theory \cite{sud1}.
This provides a platform to perform similar investigation in the   BLG theory with a Nambu-Poisson 3-bracket which describe the theory of multiple M2-branes.
 
 In this work, we review the supersymmetry in presence  of a boundary for simple
 supersymmetric theory. Further, we recapitulate the BLG theory 
 in presence of boundary condition where we show that only half of the supersymmetry   can be preserved by  adding a boundary term to the bulk. Since BLG theory on the boundary admits  a new gauge symmetry structure
based on the  3-algebra, therefore,
 to quantize such theory we choose a particular gauge. We choose DJBTM  type and Landau type gauges
in view of  their common uses. We show that the quantum actions corresponding to
these gauges admit supersymmetric BRST invariance. Further we report the existence of non-vanishing
ghost-anti-ghost condensates appeared in DJBTM gauge which leads to spontaneous supersymmetry breaking. Due to these non-vanishing ghost-anti-ghost condensates the non-linear bosonic fields
possess the non-vanishing vacuum expectation values (VEVs). The occurrence of ghost-anti-ghost condensates in MMA gauge is also shown in the  boundary BLG theory. Hence, the ghost and anti-ghost 
fields appeared in condensates are shown as the Nambu-Goldstone particles. 
We derive the effective potential for the BLG theory   which confirms
the occurrence of ghost-anti-ghost condensation and hence the spontaneous symmetry breaking in the
theory.

The paper is organized as follows. In  Sec. II, we give brief review of supersymmetry in presence of a boundary. The BLG theory in presence of boundary is studied in Sec. III.
In Sec. IV, we analyse the BRST invariance of the different gauge-fixed 
actions. The effective potential corresponding to non-linear gauge and the spontaneous breaking of BRST symmetry due to
ghost-anti-ghost condensation is presented in Sec. V. Similar investigation 
for modified modified MA is reported in Sec. VI.
In the last section  the results and future investigations are given.

\section{Supersymmetry in presence of a boundary }
In this section, we analyse the possession of half of the supersymmetry, 
 without introducing explicit boundary conditions \cite{boundary,boundary0,boundary1}.
We achieve this by adding a boundary term to the 
 bulk Lagrangian of the original theory where the supersymmetric transformations of this boundary term exactly compensates the 
 boundary piece emerged from the supersymmetric transformation of the bulk Lagrangian. 
We begin with the following simple supersymmetric Lagrangian written in terms of the $\mathcal{N} =1$  superfield
defined  in three dimensions as $
 \Phi (\theta) = p + q \theta + r \theta^2,
$
where $\theta$ is a two component Grassmann parameter, 
\begin{equation}
 \mathcal{L} = D^2 [ \Phi ]_{\theta =0},\label{a}
\end{equation}
where $D^2 = D^a D_a/2$ and $ D_{a} = \partial_{a} + (\gamma^\mu \theta)_a \partial_\mu $. 
The $\mathcal{N} = 1$ supersymmetric transformations are generated by charge $Q_a$ as follows
\begin{eqnarray}
 \delta \Phi(\theta) =  \epsilon^a Q_a\Phi(\theta),
\end{eqnarray}
where charge is defined by
 \begin{eqnarray}
Q_a = \partial_a  - (\gamma^\mu \theta)_a \partial_\mu.
\end{eqnarray}
This supersymmetric 
transformation for the component fields is thus given by   
\begin{eqnarray}
  \delta p &=&\epsilon^a q_a, \nonumber \\
  \delta q_a &=& -\epsilon_a r  + (\gamma^\mu\epsilon)_a \partial_a p, \nonumber \\
  \delta r &=& \epsilon^a (\gamma^\mu \partial_\mu)_a^b q_b.
\end{eqnarray}
We remark that the Lagrangian $ \mathcal{L}$ given by Eq. (\ref{a}) on a manifold without boundaries remains invariant under these supersymmetric 
transformations, however, in presence of a boundary (say at $x_3 =0$)  
this Lagrangian under the  supersymmetric 
transformations leads a total derivative  
\begin{equation}
 \delta \mathcal{L} = - \partial_3 ( \epsilon \gamma^3 q), 
 \end{equation}
 which leads a boundary term.
This, therefore, justifies the statement that braking of the supersymmetry occurs in presence of boundary. 
To preserve  at least a part of the supersymmetry, we can  add a boundary term to the theory, such that its supersymmetric 
transformation cancels the boundary piece generated by the supersymmetric 
transformation of the bulk Lagrangian. 
Here, we  add or 
 subtract the following boundary term to the bulk Lagrangian given in Eq. (\ref{a}):
\begin{equation}
 \mathcal{L}_b =  \partial_3 [\Phi(\theta)]_{\theta =0}. 
 \end{equation}
 to preserve the supersymmetry generated  either by  boundary supercharges
$\epsilon^- Q_+$ or $\epsilon^+ Q_-$ \cite{boundary,boundary1} by adding or subtracting $\mathcal{L}_b$ to supersymmetric Lagrangian
 $\mathcal{L}$.
 Here we note that we can preserve only half of the supersymmetry because we can’t preserve
 the supersymmetries corresponding to $\epsilon^- Q_+$ and $\epsilon^+ Q_-$ instantaneously.
 Now we define the supersymmetric Lagrangian
\begin{equation}
\mathcal{L}^\pm =\mathcal{L} \pm \mathcal{L}_b =  (D^2 \mp \partial_3) [\Phi]_{\theta =0},
\end{equation}
where $\mathcal{L}^+$ denotes the Lagrangian  which preserves the supersymmetry corresponding to 
$\epsilon^- Q_+$, however, $\mathcal{L}^-$ preserves the supersymmetry   corresponding to
$\epsilon^+ Q_-$.
The decomposition of  bulk supercharge $Q_a$  is given as
$\epsilon^a Q_a = \epsilon^+ Q_- + \epsilon^- Q_+$,  
where 
$
 Q_- = Q'_- + \theta_- \partial_3, $ and $ 
 Q_+ = Q'_+ - \theta_+\partial_3, 
$
however,   $Q'_{\pm} = \partial_{\pm} - \gamma^s \theta_{\mp} \partial_s$ are the standard supersymmetry generators for the boundary fields
where  $s$ is the index for the 
coordinates along the boundary.  

\section{ BLG Theory in presence of a boundary }
In this section we recapitulate the BLG theory in presence of a boundary. 
To preserve half of the supersymmetry  of the  BLG theory we need  to add a boundary 
term as discussed in the previous section. After adding the boundary term the BLG theory 
still remains gauge variant. To retrieve a gauge invariant BLG theory, we need
one more term except the boundary term. 
Incidentally,  
  the gauge fields in the BLG theory follow the Lie $3$-algebra. A Lie $3$-algebra is simple generalization
  of Lie algebra which describes a
vector space endowed with a trilinear product, 
$
[T^A,T^B,T^C] = f^{ABC}_D T^D,
$
where  $T^A$ are   the generators of the Lie $3$-algebra. 
 The  totally antisymmetric structure constants    satisfy the following Jacobi identity, 
$
f^{[ABC}_G f^{D]EG}_H = 0
$.
In the quantization of BLG theory it is also useful to define the following structure constant, 
$
C^{AB,CD}_{EF} = 2f^{AB[C}_{[E} \delta^{D]}_{F]} 
$. 
which is anti-symmetric in the  pair of indices   and the satisfies the following Jacobi identity:
$
C^{AB,CD}_{EF} C^{GH,EF}_{KL} + C^{GH,AB}_{EF} C^{CD,EF}_{KL} +C^{CD,GH}_{EF} C^{AB,GH}_{KL} =0 $.
 The BLG theory on manifolds without boundaries preserves $\mathcal{N}  =8$ supersymmetry. 
 However, we   perform the analysis for the BLG theory on manifolds with  boundaries which preserves  $\mathcal{N} =1 $ supersymmetry  generated by the charge $Q_a = \partial_a - (\gamma^\mu \partial_\mu \theta)_a$.  
To write the Lagrangian for  BLG theory possessing the boundary term,
 we  first define super-covariant derivatives 
for  matter fields $X_A, X_A^{\dagger}$ and the spinor field $\Gamma^a_{AB}$ as follows,
\begin{eqnarray}
  \nabla_a  X^I_A&=& D_a X^I_A -i f^{BCD}_A \Gamma_{a BC} X^{I}_{ D},\nonumber\\
 \nabla_a X^{I \dagger}_A &=& D_a X^{I \dagger}_A + if^{BCD}_A X^{I \dagger}_D \Gamma^{a BC}, \\
( \nabla_a \Gamma_b)_{AB} &=& D_a \Gamma_{b AB} + C^{CD,EF}_{AB}\Gamma_{CD a} \Gamma_{b EF},  
\end{eqnarray}
where super-derivative $D_a = \partial_a + (\gamma^\mu \partial_\mu)^b_a \theta_b$.
The Lagrangian for the BLG theory on a manifold without a boundary  is defined by
\begin{eqnarray}
 \mathcal{L} &=&\nabla^2 \left [ \frac{k}{4\pi} f^{ABCD}\Gamma^{a}_{ AB}  \Omega_{a CD}
 +  \frac{1}{4} 
 (\nabla^a X^I)^A  (\nabla_a X^{\dagger}_I)_A \right. \nonumber \\  &&\left.  -\frac{2\pi}{k}
\epsilon_{IJKL} f^{ABCD}X^I_A X^{K \dagger}_B X^{J}_C  X^{L \dagger}_D\right]_{\theta =0}, 
\end{eqnarray}
where 
\begin{eqnarray}
 \Omega_{ a AB} & = & \omega_{a AB} - \frac{1}{3}C^{CD,EF}_{AB}[\Gamma^{b CD}, \Gamma_{ab EF}] \\
 \omega_{a AB} & = & \frac{1}{2} D^b D_a \Gamma_{b AB} -i  C^{CD,EF}_{AB}[\Gamma^b_{CD} , D_b \Gamma_{a EF}] \nonumber \\ && -
 \frac{1}{3}C^{CD,EF}_{AB} C^{GH,IJ}_{EF}[ \Gamma^b_{CD} ,
\{ \Gamma_{b GH} , \Gamma_{a IJ}\} ],  \\
 \Gamma_{ab AB} & = & -\frac{i}{2} \left[ D_{(a}\Gamma_{b) AB} 
- 2 i C^{CD,EF}_{AB}\{\Gamma_{a CD}, \Gamma_{b EF}\} \right]. 
\end{eqnarray}
The  matter fields $X_A, X_A^{\dagger}$ and the spinor field $\Gamma^a_{AB}$  transform under gauge transformations as follows, 
$
    \Gamma_{a } \rightarrow  i u \, \nabla_a u^{-1}, \, 
X^{I }\rightarrow  u X^I,\,
 X^{I  \dagger} \rightarrow X^{I \dagger} u^{-1},\, 
$
where 
$X^I = X^I_A T^A, \, X^{I \dagger} =  X^{I \dagger}_AT^A, \, 
 \Gamma_{a } = \Gamma_{a AB} T^A T^B $.

Now exploiting the 
a projection operator, $(P_\pm)^b_a = (\delta^b_a \pm (\gamma^3)^b_a)/2$  
the super-covariant derivatives is projected by $\nabla_{\pm b} = (P_{\pm})^a_b \nabla_a $.
Under such  projection operator the charge $Q_a$ splits up into $Q_{\pm b} = (P_{\pm})^a_b Q_a$. 
For a boundary fixed at  $x_3$, coordinate $\mu$ splits into $\mu = (\mu, 3)$, 
and as a result only half of the supersymmetry can be preserved. 
The  matter and spinor fields on the boundary 
is represented by $X_A', {X_A'}^{\dagger}$ and ${\Gamma^a_{AB}}'$, respectively.
Furthermore, the super-covariant derivative for the   fields on boundary, is written by
$\nabla_a'$. On the boundary  superfield $v$ reads $v'$. 
The Lagrangian for the  BLG theory having fixed boundary which is invariant under 
supersymmetry generated by charge $Q_+$  is given by
\begin{eqnarray}
 \mathcal{L}_{sg}&=& - \nabla_{+}' [\mathcal{CS}(\Gamma^v) + \mathcal{M} (X^{I }, X^{\dagger I })
 +\mathcal{K}' (\Gamma', v') ]_{\theta_- =0},
\end{eqnarray}
where 
\begin{eqnarray}
\mathcal{CS} (\Gamma^v) &=&\frac{k}{4\pi} \nabla_-[f^{ABCD}\Gamma^{a}_{ AB}  \Omega_{a CD}]_{\theta_{+} =0}, 
\nonumber \\ 
\mathcal{M} ( X^I, X^{\dagger I }) &=&  \frac{1}{4} \nabla_- [
 (\nabla^a X^I)^A  (\nabla_a X^{\dagger}_I)_A ]_{\theta_{+} =0}\nonumber \\  && -\frac{2\pi}{k}\nabla_- [
\epsilon_{IJKL} f^{ABCD}X^I_A X^{K \dagger}_B X^{J}_C  X^{L \dagger}_D]_{\theta_{+} =0}, \nonumber \\
\mathcal{K}' (\Gamma', v' )&=&  -\frac{k}{2\pi} 
[ f_{ABCD}
({v}'^{-1} \nabla_{+}' v')^{AB} ({{v}'}^{-1} \mathcal{D}_{-}' v')^{CD}],
\end{eqnarray}
$\Gamma^v_a$ denotes the gauge transformation  of $\Gamma_a$ generated by $v$.

It may be noted that the difference 
 $ \mathcal{CS}(\Gamma^v) - \mathcal{CS}(\Gamma)=\mathcal{S}(\Gamma', v')$ defines the boundary potential.  So, the total
potential of the theory reads
$\mathcal{CS}(\Gamma^v) =\mathcal{CS}(\Gamma) +  \mathcal{S}(\Gamma', v') ) $.
 In the absence of coupling between the gauge and the bulk fields this reduces to a potential term
 for the supersymmetric Wess-Zumino-Witten models 
\begin{eqnarray}
\nabla_{+}' \mathcal{S}(\Gamma', v') &=& -\frac{k}{2\pi} {\nabla}_{+} 'C^{CD, EF}_{AB}
\left[ [({v^{-1}}' \mathcal{D}_{-}' v')^{AB},  
({v^{-1}}' \mathcal{D}'_3 v')_{CD}]\right. \nonumber \\ && \left. \times
({v^{-1}}' \nabla_{+}' v')_{EF} \right]_{\theta_- =0}.
\end{eqnarray}
\section{The gauge-fixed boundary BLG theory} 
The gauge invariance of the BLG theory on the boundary ponders the presence of
spurious degrees of freedom in the theory. Consequently, we cannot quantize it without fixing a gauge.
 In this case, therefore, we make the following choice of gauge fixing condition, 
\begin{eqnarray}
G \equiv  D^a  \Gamma_{aAB} =0.\label{gf}
\end{eqnarray}

Here if  the bulk fields respect some boundary conditions then
the   path integral must be a sum over the bulk fields obeying those boundary conditions.
However, by including the boundary fields in the path integral, both the bulk  and
and the boundary fields will be integrated over.  Henceforth, the
bulk fields can be split into a pure bulk component and a boundary component. 
This can be achieved  by first including  the separate boundary fields and then introducing the 
Lagrange multipliers to constrain those boundary fields to match the boundary limits of the bulk fields. 
Therefore, we need separate  gauge-fixing terms for  he bulk and   boundary fields in the BLG theory on a 
boundary. 
The boundary gauge-fixing condition can be constructed in such a way that it will lead   the boundary limit of the bulk gauge-fixing condition.
Keeping these points in mind,  the  gauge-fixing condition (\ref{gf}) can be incorporated 
in the BLG theory at quantum level  by adding following term in the 
invariant Lagrangian:
\begin{equation}
\mathcal{L}_{gf} =\nabla_+ \nabla_- 
 \left[f^{ABCD}b_{AB}  D^a \Gamma_{a CD} + \frac{\alpha}{2}f^{ABCD} b_{AB}  b_{CD}  
\right]_{\theta =0}.
\end{equation}
The induced Faddeev-Popov ghost term  for this gauge fixing term is written by 
\begin{equation}
\mathcal{L}_{gh} = i\nabla_+ \nabla_- 
\left[ f^{ABCD}\bar{c}_{AB}   D^a \nabla_a   c_{CD}
\right]_{\theta =0}.
\end{equation}
 Incorporating these terms the total Lagrangian density for the boundary BLG theory in 
 Lorentz type gauge  is given by  
\begin{eqnarray}
  \mathcal {L}_{BLG}= 
 \mathcal {L}_{sg}+\mathcal{L}_{gf} +\mathcal{L}_{gh}.
\end{eqnarray}
This  Lagrangian density for BLG theory on boundary remains invariant under 
the following   BRST transformations: 
\begin{eqnarray}
&&s \,\Gamma_{aAB} = \nabla_a   c_{AB},\ 
s \,c_{AB} = - \frac{1}{2} C^{EF,GH}_{AB}c_{EF} c_{GH} \   ,
\nonumber \\
&&s \,\bar{c}_{AB} = ib_{AB}, \
s \, X_A^{ I \dagger }
 =  - i  X^{I B\dagger } c_{AB},  \nonumber \\ &&  s \,b_{AB} =0, \
s \, X^{I  }_A = ic_{AB}  X^{IB }\   , 
  \nonumber \\
 && s\, v_{AB} = -i  C^{EF,GH}_{AB}v_{EF} c_{GH}.  \label{brs}
\end{eqnarray}
This is also invariant under the another set of supersymmetric transformations called as the anti-BRST transformations which are given by 
\begin{eqnarray}
 && \bar s \,\Gamma_{aAB} = \nabla_a   \bar c_{AB}, \
\bar s \,\bar c = - \frac{1}{2} \frac{1}{2} C^{EF,GH}_{AB}\bar c_{EF}\bar c_{GH},
\nonumber \\
 && \bar s \, X_A^{ I \dagger }
 =  - i  X^{I B\dagger } \bar c_{AB}, \
\bar s\, \bar b_{AB} =0,  \nonumber \\ 
 &&\bar s \, X^{I  }_A=i\bar c_{AB}  X^{I B},  \ 
\bar s\,{c}_{AB} =i\bar b_{AB},
   \nonumber \\
 &&  \bar s\, v_{AB} = -i   C^{EF,GH}_{AB}v_{EF} \bar{ c}_{GH},\label{abrs}
\end{eqnarray}
where new auxiliary field is expressed in terms of original fields as follows,
\begin{equation}
\bar b_{AB}=-b_{AB}+i  C^{EF,GH}_{AB}c_{EF} \bar c_{GH}.\label{cf}
\end{equation}  
This defines the Curci-Ferrari (CF) type restriction.

In the particular limit, $\alpha =0$, the Lorentz gauge corresponds to the  Landau gauge, 
and in this scenario the sum of the gauge fixing  and the 
ghost terms can be written as both the  BRST and anti-BRST exact terms
\begin{eqnarray}
\mathcal{L}_{gf} + \mathcal{L}_{gh} &=&  \frac{i}{2}\nabla_+ \nabla_- 
s\bar s \left[ f^{ABCD}\Gamma^{a}_{AB}\Gamma_{a CD}
\right]_{\theta =0}, \nonumber \\ 
&=& -\frac{i}{2}\nabla_+ \nabla_- 
\bar s s \left[ f^{ABCD}\Gamma^{a}_{AB}\Gamma_{a CD}
\right]_{\theta =0}.
\end{eqnarray}
 Here we note that the BRST or the anti-BRST transformations of the original BLG theory produce a surface
 term which is compensated by the BRST or the anti-BRST variations of the   boundary  term of 
 the modified BLG theory. In this way, 
  the overall BRST and the anti-BRST invariances of the BLG theory on boundary  are recovered.

Now,  we analyse  the BLG theory on boundary 
in   DJBTM (non-linear)  gauge.
 In the DJBTM  gauge, the sum of 
the  gauge-fixing and ghost terms of the effective Lagrangian    
 is given by
\begin{eqnarray}
{\cal L}^{DJ}_g 
&=&\nabla_+ \nabla_-    \left[f^{ABCD} b_{AB}  D^a \Gamma_{aCD} + \frac{\alpha}{2}f^{ABCD}b_{AB}b_{CD} 
  + if^{ABCD}\bar{c}_{AB}    D^a \nabla_a  c_{CD}\right.\nonumber\\ 
  &&\left.   -i\frac{\alpha}{2}  C^{EF,GH}_{AB}  c_{EF}\bar c_{GH}b^{AB} +\frac{\alpha}{8} C^{EF,GH}_{AB} C_{IJ,KL}^{AB} \bar c_{EF} \bar c_{GH}c^{IJ} c^{KL}
  \right]_{\theta =0}. 
 \end{eqnarray}
Further,  it can be expressed by
 \begin{eqnarray}
{\cal L}^{DJ}_g 
 &=&\nabla_+ \nabla_-   \left[ f^{ABCD} b_{AB}  D^a \Gamma_{aCD} + \frac{\alpha}{2}f^{ABCD}b_{AB}b_{CD} 
  + if^{ABCD}\bar{c}_{AB}    D^a \nabla_a  c_{CD}\right.\nonumber\\ 
  &&\left.   -i\frac{\alpha}{2}  C^{EF,GH}_{AB}  c_{EF}\bar c_{GH}b^{AB} -\frac{\alpha}{4}C^{EF,GH}_{AB} C_{IJ,KL}^{AB}  c_{EF} \bar c_{GH}c^{IJ} \bar c^{KL}
  \right]_{\theta =0}.
\end{eqnarray}
To scrutinize the non-zero gauge 
parameter, it is expressed by
\begin{eqnarray}
{\cal L}^{DJ}_g &=&\nabla_+ \nabla_-  \left[ \frac{\alpha}{2}\left(b_{AB}-\frac{i}{2}C^{EF,GH}_{AB}  c_{EF} \bar c_{GH}  +\frac{1}{\alpha} D_a\Gamma^a_{AB} \right)^2 -\frac{1}{2\alpha}(D_a\Gamma^a_{AB})^2 
  \right.\nonumber\\
    &&\left.  + if^{ABCD}\bar{c}_{AB}    D^a \nabla_a    c_{CD}  -\frac{\alpha}{8}C^{EF,GH}_{AB} C_{IJ,KL}^{AB}  c_{EF} \bar c_{GH}c^{IJ} \bar c^{KL}
  \right]_{\theta =0}.
\end{eqnarray}
The total Lagrangian  density for BLG theory on boundary  in DJBTM  gauge is written as the sum of invariant
part and the gauge-fixed part,
\begin{equation}
{\cal L}'_{BLG}= {\cal L}_{sg} + {\cal L}^{DJ}_{g}, 
\end{equation}
which remains 
invariant under the BRST and anti-BRST transformations given respectively
  in   (\ref{brs}) and (\ref{abrs}).
The gauge-fixed part of the above Lagrangian density   ${\cal L}^{DJ}_g$ can be expressed as the
 BRST and anti-BRST exact terms as follows, 
\begin{eqnarray}
{\cal L}^{DJ}_g  &=& \frac{i}{2}\nabla_+ \nabla_- 
s\bar s \left[ f^{ABCD}(\Gamma^{a}_{AB}\Gamma_{a CD}
- \alpha \bar c_{AB} c_{CD})\right]_{\theta =0}, \nonumber \\
 &=& -\frac{i}{2}\nabla_+ \nabla_- 
\bar s s \left[ f^{ABCD}(\Gamma^{a}_{AB}\Gamma_{a CD}- \alpha \bar c_{AB} c_{CD})
\right]_{\theta =0}.
\end{eqnarray}
The non-linear auxiliary 
field $b$   plays an important 
role as an order
parameters  in analysing the  spontaneous breaking of BRST symmetry.  
\section{Effective potential for boundary BLG theory in non-linear gauge}
In this section, we investigate  the spontaneous breakdown of  BRST 
supersymmetry  
in boundary BLG theory.
In this context, we first
 define the potential $V(b )$ for multiplier
 fields 
 $b_{AB}$   as follows
 \begin{eqnarray}
 V(b ) &=& \nabla_+ \nabla_-  \left[- \frac{\alpha}{2}\left(b_{AB}-\frac{i}{2}C^{EF,GH}_{AB}  c_{EF} \bar 
 c_{GH}  +\frac{1}{\alpha} D_a\Gamma^a_{AB} \right)^2 \right]_{\theta =0}. 
 \end{eqnarray}
From this expression it is evident that the potential has  extremum for gauge parameter $\alpha$  at
\begin{eqnarray}
b_{AB}=\frac{i}{2}C^{EF,GH}_{AB}  c_{EF} \bar c_{GH}  -\frac{1}{\alpha} D_a\Gamma^a_{AB}.
\end{eqnarray}
The  vacuum expectation value
of non-linear bosonic field  $b$ corresponding to the vanishing spinor expectation value
 $\langle \Gamma^a_{AB}\rangle =0$  takes
\begin{equation}
\langle 0| b_{AB}|0\rangle =\frac{1}{2}\langle 0|iC^{EF,GH}_{AB}  c_{EF} \bar c_{GH} |0\rangle.
\end{equation}
  In presence of ghost-anti-ghost condensation 
\begin{eqnarray}
   \langle 0|iC^{EF,GH}_{AB}  c_{EF} \bar c_{GH} |0\rangle \neq 0, 
\end{eqnarray}
 the non-linear field  $b_{AB}$ 
acquires the non-vanishing vacuum   expectation value (VEV), i.e., 
\begin{eqnarray}
\langle 0| b_{AB}|0\rangle =\frac{1}{2}\langle 0|iC^{EF,GH}_{AB}  c_{EF} \bar c_{GH} |0\rangle \neq 0.
\end{eqnarray}
As a result,   the spontaneous breaking in BRST symmetry occurs due to this non-vanishing VEV,  
\begin{eqnarray}
\langle 0| s  \bar c_{AB}|0\rangle = \langle 0|i b_{AB}|0\rangle =-\frac{1}{2}\langle 0| C^{EF,GH}_{AB}  c_{EF} \bar c_{GH} |0\rangle \neq 0.
\end{eqnarray}
Exploiting CF type condition (\ref{cf}), we observe that
  the spontaneous breaking of anti-BRST symmetry also occurs,
\begin{eqnarray}
\langle 0|\bar s  c_{AB}|0\rangle = \langle 0|i \bar b_{AB}|0\rangle =-\frac{1}{2}\langle 0|C^{EF,GH}_{AB}  c_{EF} \bar c_{GH} |0\rangle \neq 0.
\end{eqnarray}
The  spontaneous breaking of BRST and anti-BRST symmetries reflect the presence of massless  Nambu-Goldstone particles for boundary BLG theory  following  Nambu-Goldstone theorem. 
Here the  ghosts and anti-ghosts are Nambu-Goldstone particles.

To specify whether such ghost-anti-ghost condensations and therefore
spontaneous symmetry breaking take place or not,
it is important
to evaluate the effective potential for the composite operator $i C^{EF,GH}_{AB}  c_{EF} \bar c_{GH} $.
Performing the analysis for Lie 3-algebra as in \cite{kon} leads to the total
bosonic effective potential 
\begin{eqnarray}
V(b,  \phi ) =
V(\phi )+\nabla_+ \nabla_-  \left[ -\frac{\alpha}{2} \left( b_{AB} +\frac{1}{2\alpha} \phi_{AB}\right)^2
\right]_{\theta =0},
\end{eqnarray} 
where $V(\phi )$ refers  the
effective potential  for $\phi_{AB} \sim -\alpha \langle 0|i C^{EF,GH}_{AB}  c_{EF} \bar c_{GH}  |0\rangle$.
Now we see that the potential $V(\phi)$ has stationary points for
 $b_{AB}=-\frac{1}{2\alpha} \phi_{AB}$. Hence the 
condensation is meaningful for $\alpha>0$ only. 
However,   
for Landau gauge condition ($\alpha =0$) where gauge parameter takes zero value
the minimum or maximum of the potential of the field $b_{AB}$ occurs at $b_{AB}=0$,
which implies $\langle 0|i \bar b_{AB}|0\rangle =0 $. Therefore, in the Landau gauge the spontaneous 
BRST symmetry breaking due to the mechanism mentioned above can not occur (at least in
the tree level).

\section{Effective potential for boundary BLG theory in MMA gauge  }
In this section, we discuss ghost-anti-ghost condensation for boundary BLG theory 
in  MMA  gauge  \cite{4z, 5z}. To do so, let us begin  by 
decomposing  the spinor field  in diagonal and off-diagonal components as
follows
 \begin{eqnarray}
 \Gamma^a_{AB}= \gamma_{aAB}^iT_i +  {\Upsilon}^\alpha_{aAB} T_\alpha,
\end{eqnarray}
where $T_i\in \mathcal{\cal H}$ and $T_\alpha\in \mathcal{\cal G}-\mathcal{\cal H}$.
Here $\mathcal{\cal H}$ refers to the Cartan subalgebra of the Lie algebra $\mathcal{\cal G}$.
Now, the gauge-fixed Lagrangian  density  in MA gauge incorporating    diagonal and off-diagonal 
decomposition  is  given by
\begin{eqnarray}
{\cal L}_g^{MA} &=&-i\nabla_+ \nabla_-  s \left[f^{ABCD}\bar c_{AB}\left\lbrace \nabla_a[\gamma]\Upsilon^a_{CD} +\frac{\alpha}{2}b_{AB}\right\rbrace -i\frac{\zeta}{2}C^{EF,GH}_{AB} \bar c^{AB} \bar{c}_{EF} c_{GH} \right.
\nonumber\\
&-&\left. i\frac{\zeta}{4}C^{EF,GH}_{AB} c^{AB} \bar{c}_{EF}  \bar c_{GH} \right]_{\theta =0},
\end{eqnarray}
which can further be expanded by assigning the  BRST transformation on the superfields as follows
\begin{eqnarray}
{\cal L}_g^{MA} &=&\nabla_+ \nabla_-  \left[f^{ABCD}b_{AB}\nabla_a [\gamma]\Upsilon^a_{CD}
+ \frac{\alpha}{2}f^{ABCD}b_{AB}b_{CD}\right.\nonumber\\
&&\left. +if^{ABCD}\bar c_{AB}\nabla_a[\gamma]\nabla^a [\gamma]c_{CD}-iC^{EF,GH}_{AB}C_{IJ,KL}^{AB}  \bar c_{EF} \Upsilon_{aGH}  c^{IJ}\Upsilon^{aKL} \right.\nonumber\\
&&\left.+iC^{EF,GH}_{AB} \bar c^{AB}\nabla_a[\gamma]   (\Upsilon^a_{EF} c_{GH} )+ iC^{EF,GH}_{AB} \bar c^{AB}\nabla_a[\gamma]\Upsilon^a_{EF} c_{GH} \right.\nonumber\\
&&\left.+\frac{\zeta}{8}C^{EF,GH}_{AB}C_{IJ,KL}^{AB}  \bar c_{EF} \bar c_{GH} c^{IJ}c^{KL} +\frac{\zeta}{4}C^{EF,GH}_{AB}C_{IJ,KL}^{AB} \bar c_{EF} \bar c_{GH} c^{IJ}c^{KL}
\right.\nonumber\\
&&\left. +i\frac{\zeta}{2}C^{EF,GH}_{AB}c^{AB}b_{EF}\bar c_{GH}-i\zeta C^{EF,GH}_{AB} b^{AB}\bar c_{EF}c_{GH} \right.\nonumber\\
&&\left. +\frac{\zeta}{4}C^{EF,GH}_{AB}C_{IJ,KL}^{AB} \bar c_{EF} \bar c_{GH} c^{IJ}c^{KL}
\right]_{\theta =0}.
\end{eqnarray}
The requirement of the orthosymplectic 
invariance of the MMA gauge yields the quartic ghost interaction as $\zeta =\alpha$. Therefore,  the above expression in the   MMA gauge  reduces to
\begin{eqnarray}
{\cal L}_g^{MMA} &=&\nabla_+ \nabla_-  \left[\frac{\alpha}{2}\left(b_{AB}- iC^{EF,GH}_{AB}\bar c_{EF} c_{GH} 
+\frac{1}{\alpha} \nabla_a[\gamma] \Upsilon^a_{AB}\right)^2  -\frac{1}{2\alpha} (\nabla_a[\gamma] \Upsilon^a_{AB} )^2 
\right.\nonumber\\
&&\left. - iC^{EF,GH}_{AB}C_{IJ,KL}^{AB} \bar c_{EF} c_{GH}   \Upsilon^{IJ}_a \Upsilon^{aKL}  + iC^{EF,GH}_{AB}\bar c^{AB}\nabla_a[\gamma]( \Upsilon^a_{EF} c_{GH} ) 
  \right.\nonumber\\
&&\left.  -if^{ABCD}\bar c_{AB}\nabla_a[\gamma]\nabla^a [\gamma]c_{CD} 
\right]_{\theta =0}.
\end{eqnarray}
For the MMA gauge, the potential for non-linear field $b$   has its extremum at
\begin{eqnarray}
b_{AB}&=& iC^{EF,GH}_{AB}\bar c_{EF} c_{GH} -\frac{1}{\alpha} \nabla_a[\gamma] \Upsilon^a_{AB}.
\end{eqnarray}
So, the  VEV in this case reads 
\begin{eqnarray}
\langle 0|b_{AB}|0\rangle &=& \langle 0|iC^{EF,GH}_{AB}\bar c_{EF} c_{GH}|0\rangle -\frac{1}{\alpha} \langle 0| \nabla_a[\gamma] \Upsilon^a_{AB}|0\rangle.
\end{eqnarray}
One can obtain gauge parameter $\beta$ dependence of the
vacuum-to-vacuum amplitude $Z$ as following:
\begin{eqnarray}
\frac{\delta Z}{\delta\beta} =\frac{1}{2}\int d^3 x \langle 0; \mbox{out}|s\left(\bar c_{AB} (x)b^{AB}(x)\right)|0;\mbox{in}\rangle
\end{eqnarray}
It signifies that  the BLG theories with different gauge parameters are different theories.
In this case 
the total bosonic effective potential
is computed  by
\begin{eqnarray}
V(\bar b, b, \phi)=V(\phi)-\frac{\beta}{2}\left(\bar b+\frac{1}{\zeta} \phi  \right)^2-\frac{\alpha}{2}
b_{AB}b^{AB},
\end{eqnarray}
where we have omitted the index of the diagonal component.
Here we note that the effective potential $V(\phi)$  has minima at non-zero values of $\alpha$.
Therefore,
the spontaneous breakdown of the BRST or anti-BRST  could happen, if $\zeta>0$ and $\beta<0$.
Since the total bosonic effective potential has an absolute minimum
at non-zero value of $\bar b=-\frac{1}{\zeta} \phi$.
This shows that  due to presence of the ghost-anti-ghost condensates the boundary BLG
 theory in MMA gauge  
the spontaneous breakdown of the BRST symmetry occurs.

The spontaneous BRST and anti-BRST supersymmetry breaking reflect that
the Nambu-Goldstone particles  associated with these can be  identified as the 
diagonal anti-ghost  or diagonal ghost, respectively. It means that the 
diagonal ghost and the diagonal anti-ghost are massless particles which are consistent with the
infrared Abelian dominance. Since for the  infrared Abelian dominance,  
the off-diagonal components of ghosts become 
massive while the diagonal components remain massless. 

\section{Conclusion}
As we know, the basic objects to unify  string theories in ten dimensions are M2-branes and M5-branes
(the magnetic version of M2-branes in the sence that M5-branes couple to the
dual background three form C-field in 11D supergravity). Therefore we could consider
that M-branes are the most fundamental objects \cite{bl}. To understand the
mysterious nature of the M-theory, it is desirable to understand properties of
 M2-branes and M5-branes. A single M2-brane or a
single M5-brane had already been known for quite a long time.  However, 
the   multiple M2 branes is studied  in recent past  years by BLG
 theory and ABJM theory \cite{bl0,bl00, bl01,bl02,bl1}.
 In fact, the BLG theory has been 
 identified with the M5-brane action in presence of a large  C-field \cite{ber4}.  
It is obvious that BLG theory by itself cannot be identified with a  6 dimensional theory, as it is a 3 dimensional theory.
 
In this paper, we analysed the BLG theory which follows the Lie 3-algebra in different gauges.
In view of
their extreme importance, we choose these gauges to be the  Delbourgo-Jarvis and Baulieu-Thierry-Mieg (DJBTM) gauge  and 
modified maximally Abelian (MMA) gauge.  
The quantum actions corresponding to these gauges admit supersymmetric
BRST invariance.
We have shown that the
existence of non-vanishing  ghost-anti-ghost condensates appeared in  DJBTM gauge which can also be
justified by symmetry breaking considerations. 
Due to  these  non-vanishing  ghost-anti-ghost condensates
the non-linear bosonic fields possess the non-vanishing 
vacuum expectation values (VEVs). 
The occurrence of ghost-anti-ghost condensates  in MMA gauge is also found in the BLG theory which
causes the  spontaneous breaking of the BRST symmetry.
The  ghost  and anti-ghost fields involved in the condensates   are identified as
 Nambu-Goldstone particles. The expression for effective potential for the BLG theory is
 given which confirms the occurrence  of ghost-anti-ghost condensation 
and hence the spontaneous symmetry breaking in the theory.
Our present investigation will help in performing the numeric simulations for the propagator for 
non-linear gauge.
Also this analysis will help in understanding the more explicit examples of Lie 3-algebra.
Another possible extension of the present work is  to explore this in an alternative formalism for boundary supersymmetry involving SIM (1) superspace \cite{v,v1,v2}.

\end{document}